\documentclass[12pt]{article}
\usepackage{amssymb}
\begin{document}

\vskip2cm

\begin{center}
{\bf THE THREE-NEUTRINO MIXING AND OSCILLATIONS}

\vspace{0.3cm} S. M. Bilenky
\footnote {Lectures at Corfu Summer Institute on Elementary Particle Physics,
Corfu, 31.8 - 20.9. 2001 }\\
\vspace{0.3cm} {\em IFAE, Facultat de Ciencies, Universidad Autonoma
de Barcelona, 08193, Bellaterra, Barcelona, Spain \\}

\vspace{0.3cm} {\em  Joint Institute
for Nuclear Research, Dubna, R-141980, Russia\\}
\end{center}

\begin{abstract}
The basics of neutrino oscillations is presented. Existing
evidences of neutrino oscillations, obtained in the atmospheric
and solar neutrino experiments, are considered. The new CHOOZ
bound on the element $|U_{e3}|^{2}$, obtained from the
three-neutrino analysis of the data, is discussed. Decoupling of
neutrino oscillations in the solar neutrino range of $\Delta
m^{2}$ is considered.
\end{abstract}

\section {Introduction}

There exist at present strong model independent evidences in favor
of neutrino oscillations, obtained in the atmospheric
\cite{AS-K,Soudan,MACRO} and in the solar
\cite{Cl,Kam,GALLEX,GNO,SAGE,S-K,SNO} neutrino experiments. The
direct evidence for oscillations of atmospheric neutrinos is the
significant up-down asymmetry of the high-energy muon events,
observed in the atmospheric Super-Kamiokande (S-K) experiment
\cite{AS-K}. The three-sigma proof of the presence of $\nu_{\mu}$
and $\nu_{\tau}$ in the flux of the solar neutrinos on the earth
that stems from the comparison of the results of the SNO
\cite{SNO} and S-K \cite{S-K} solar neutrino experiments is the
direct evidence for the oscillation of solar neutrinos.

Indications in favor of
$\bar \nu_{\mu}\to \bar\nu_{e} $ oscillations were obtained in the
accelerator LSND experiment \cite{LSND} . This result requires confirmation
and it will be checked by the MiniBooNE experiment
 \cite{MiniB}, started recently.
We will not consider LSND result here.

The data of the S-K atmospheric neutrino experiment are perfectly
described if we assume that the two-neutrino
$\nu_{\mu}\to\nu_{\tau} $ oscillations take place. From the
analysis of the data of the S-K experiment the following best-fit
values for the neutrino oscillation parameters were found
\cite{AS-K}

\begin{equation}
\Delta m^{2}_{\rm{atm}}=2.5 \cdot 10^{-3}\,\rm{eV}^{2};\,~~
\sin^{2}2\theta_{\rm{atm}} =1 \,.
\label{001}
\end{equation}

The data of all solar neutrino experiments can be described if we
assume that the probability of solar $\nu_{e}$ to survive
has the two-neutrino form is
characterized by two parameters $\Delta m^{2}_{sol}$ and
$\tan^{2}\theta_{sol}$. From
the global analysis of all solar neutrino data several allowed
regions in the plane of the oscillation parameters were found
 \cite{Os,Fogli,Bahcall,Band}.
For the most favorable LMA MSW region the best-fit values of the parameters
 are \cite{Bahcall}
\begin{equation}
 \Delta m^{2}_{\rm{sol}} = 3.7 \cdot 10^{-5}\,\rm{eV}^{2};\,~~
 \tan^{2}\theta_{\rm{sol}} = 3.7 \cdot 10^{-1}\,.
\label{002}
\end{equation}

Let us note that for other allowed regions the values of
 $\Delta m^{2}_{sol}$ are significantly smaller than for the LMA region.
Thus, from  the analysis of the existing neutrino oscillation data
it follows that neutrino mass squared differences,
relevant for the oscillations of the solar and atmospheric neutrinos,
satisfy
the hierarchy relation

\begin{equation}
 \Delta m^{2}_{\rm{sol}}\ll\Delta m^{2}_{\rm{atm}}\,.
\label{003}
\end{equation}

\section{Neutrino mixing}

Neutrino oscillations are driven by the
neutrino mixing
\begin{equation}
\nu_{\alpha L}
=
\sum_{i=1}^3 U_{\alpha i} \nu_{iL}\,, \qquad ({\alpha}=e,\mu,\tau)
\label{004}
\end{equation}

where $\nu_{\alpha L}$ is the flavor neutrino field, $\nu_{i}$ is
the field of neutrino (Dirac or Majorana) with mass $m_{i}$ and $U$
is the unitary mixing matrix.

From the data of the SLC and LEP experiments on the measurement of
the width of the decay of the $Z$-boson into neutrino-antineutrino
pairs it follows that only three flavor neutrinos exist in nature. In
the latest LEP experiments, for the number of the flavor neutrinos
the value \cite{PDG}

$$n_{\nu_{f}} = 3.00 \pm 0.06 $$
was obtained.

 The minimal number of massive neutrinos is equal to
the number
of flavor neutrinos (three).
If the number of massive neutrinos $n$
is larger than three,
in this case $n - 3$ sterile neutrinos must exist
(see, for example, \cite{BGG}).

The  data of the solar and atmospheric neutrino experiments can be described
in the framework of the {\em minimal} scheme with the number of
massive neutrinos being equal to the measured number of flavor
neutrinos. Let us note, that if LSND data will be confirmed, it
would mean that a third independent mass-squared difference must
exist and the minimal number of massive neutrinos must be equal to
four.

From (\ref{004})
for the state of the flavor neutrino with momentum $\vec p$ we have

\begin{equation}
|\nu_\alpha\rangle
=
\sum_{i} U_{{\alpha}i}^* \,~ |\nu_i\rangle
\,,
\label{005}
\end{equation}
where $|\nu_i\rangle$ is the state of neutrino with momentum
$\vec p$ and energy
$E_i
=
\sqrt{p^2 + m_i^2 }
\simeq
p + \frac{ m_i^2 }{ 2 p }$.

The relation (\ref{005}) is the basic one. We will make a few relevant comments. Let us consider a decay
\begin{equation}
a \to b + l^{+}+ \nu_{l}\,,
\label{006}
\end{equation}
in which together with a lepton $l^{+}$ a flavor neutrino
$\nu_{l}$ is produced. For the state of neutrinos with momentum
$\vec p$ we have from relation (\ref{004})

\begin{equation}
 |\nu>_{\vec p} = \,~~\sum_{i}|\nu_i\rangle \,~ \langle i\,~b\,~l^{+}~| S |\,~a\rangle \,,
\label{007}
\end{equation}
where $|\nu_i\rangle$ is the state of neutrino with momentum
$\vec p$ and energy $E_i$.

Taking into account that neutrino mass-squared differences $\Delta m^{2}$
are much smaller
than the square of neutrino energy
($\Delta m^{2}/E^{2}\lesssim 10^{-15}$), we have \cite{BilG}
\begin{equation}
\langle i\,~b\,~l^{+}~| S |\,~a\rangle\simeq U_{li}^{*}\,~
\langle \nu_{l}\,~b\,~l^{+}~| S |\,~a\rangle_{SM}\,.
\label{008}
\end{equation}
Here $\langle \nu_{l}\,~b\,~l^{+}~| S |\,~a\rangle_{SM}$ is the
Standard Model matrix element of the process (\ref{006}),
calculated under the assumption that the mass of $\nu_{l}$ is
equal to the minimal neutrino mass $m_{1}$. This matrix element
does not depend on $i$. From Eq.(\ref{007}) and Eq.(\ref{008}) we
obtain relation (\ref{005}) for the normalized state of the flavor
neutrino. \footnote{ Due to the uncertainty relation, the emission
(absorption) of neutrinos with different masses and very small
mass-squared differences can not be resolved. This is the reason,
why flavor neutrinos are described by the {\em coherent
superposition} of states of neutrinos with definite masses (see
\cite{GrS}).}

We will consider next the standard parameterization of the
neutrino mixing matrix $U$. If $\nu_{i}$ are Dirac particles, the
unitary n $\times$ n mixing matrix is characterized by
$n_{\theta}= n (n-1)/2$ angles and $n_{\phi}= (n-1)(n-2)/2$
phases. If $\nu_{i}$ are Majorana particles, the mixing matrix is
characterized by $n_{\theta}= n (n-1)/2$ angles and $n_{\phi}=
n(n-1)/2$ phases. The additional $n-1$ phases in the Majorana case
do not enter, however,
 into the expressions for the neutrino transition probabilities \cite{BHP,Doi}.
Thus, in the case of three massive neutrinos the neutrino mixing
matrix in the expression for the neutrino transition probability
both in the Dirac and in the Majorana cases is characterized by
three mixing angles and one phase.

In order to parameterize the 3$\times$3 PMNS
(Pontecorvo-Maki-Nakagawa-Sakata) mixing matrix \footnote{
B.Pontecorvo considered neutrino oscillations in 1958 after the
$V-A$ theory of the weak interaction was proposed. At that time
only one type of neutrino was known. In his first pioneering paper
\cite{Pon}
 B.Pontecorvo discussed transitions of active neutrino
into sterile state. He considered neutrino oscillations as a
phenomenon analogous to the oscillations of neutral kaons. Later,
after the second neutrino was discovered, it was not difficult for
him \cite{BPont} to generalize the idea of oscillations for the
case of two flavor neutrinos \cite{BPon}. Maki, Nakagawa and
Sakata \cite{MNS} in 1962 proposed the mixing of two massive
neutrinos.} we will construct three vectors $|\nu_{e}\rangle$,
$|\nu_{\mu}\rangle$ and $|\nu_{\tau}\rangle$, that satisfy the
condition
\begin{equation}
\langle\nu_{\alpha'}|\nu_{\alpha}\rangle = \delta_{\alpha';\alpha}\,.
\label{009}
\end{equation}
Let us start with $|\nu_{e}\rangle$. Taking into account that

$$\sum_{i=1,2} |U_{e i}|^{2} = 1 -|U_{e 3}|^{2}$$
we will introduce the angle $\theta_{12}$ in
 the following way
\begin{equation}
U_{e 1}= \sqrt{ 1 -|U_{e 3}|^{2}}\,~\cos\theta_{12}\,;~~
U_{e 2}= \sqrt{ 1 -|U_{e 3}|^{2}}\,~\sin\theta_{12}\,.
\label{010}
\end{equation}
For the vector  $\nu_{e}$ we have
\begin{equation}
 |\nu_{e}\rangle=\sqrt{ 1 -|U_{e 3}|^{2}}\,~|n_{12}\rangle + U_{e 3}\,~|3\rangle\,,
\label{011}
\end{equation}
where

$$|n_{12}\rangle = \cos\theta_{12}\,~|1\rangle + \sin\theta_{12}\,~|2\rangle\,.$$

The CP violating phase $\delta$
is introduced as follows
\begin{equation}
U_{e 3} = |U_{e 3}|\,~ exp{(-i \delta)}\,.
\label{012}
\end{equation}

Two unit vectors

$$|n_{12}^{\bot}\rangle = -\sin\theta_{12}\,~|1\rangle +
\cos\theta_{12}\,~|2\rangle$$

and
$$
|n_{3}\rangle =U_{e 3}^{*}\,~|n_{12}\rangle + \sqrt{ 1 -|U_{e 3}|^{2}}\,~|3\rangle$$

are orthogonal to $ |\nu_{e}\rangle$. We will introduce now the angle
$\theta_{23}$ in the following way
\begin{equation}
|\nu_{\mu}\rangle = \cos\theta_{23}\,~|n_{12}^{\bot}\rangle +  \sin\theta_{23}\,~|n_{3}\rangle
\label{013}
\end{equation}

and

\begin{equation}
|\nu_{\tau}\rangle =- \sin\theta_{23}\,~|n_{12}^{\bot}\rangle +  \cos\theta_{23}\,~|n_{3}\rangle\,.
\label{014}
\end{equation}

Using (\ref{011}), (\ref{013}) and (\ref{014}), we can easily
express all elements of the mixing matrix through the mixing
angles and the CP-violating phase \cite{PDG}. For the elements of
the first row and the third column we have
\begin{equation}
U_{e 1} = \sqrt{ 1 -|U_{e 3}|^{2}}\,~ \cos\theta_{12}\,;
U_{e 2} = \sqrt{ 1 -|U_{e 3}|^{2}}\,~ \sin\theta_{12}\,.
\label{015}
\end{equation}

and
\begin{equation}
U_{\mu 3} = \sqrt{ 1 -|U_{e 3}|^{2}}\,~ \sin\theta_{23}\,;
U_{\tau 3} = \sqrt{ 1 -|U_{e 3}|^{2}}\,~ \cos\theta_{23}\,.
\label{016}
\end{equation}

From the analysis of the atmospheric neutrino data it follows that
the angle $\theta_{23} \simeq \theta_{\rm{atm}} $ is large (close
to $\pi /4$). In the case of the most plausible LMA MSW fit of the
solar neutrino data the angle
 $\theta_{12} \simeq \theta_{\rm{sol}}$ is also large. As we will see later,
from the analysis of the data of the long baseline reactor
experiments CHOOZ \cite{CHOOZ} and Palo Verde \cite{PaloV} it follows that the element $|U_{e 3}|^{2}$ is small.

\section{Neutrino oscillations}

Let us now turn to  neutrino oscillations. From the basic relation
(\ref{005}) we obtain for the amplitude of the transition
$\nu_\alpha \rightarrow \nu_{\alpha'}$ in vacuum the following
relation (see, for example, \cite{BP,BPet})
\begin{equation}
 A_{\nu_{\alpha} \to \nu_{\alpha'}}(t)=\langle\nu_{\alpha'}|
\,~ \exp{(-iH_{0}t)}\,~|\nu_{\alpha}\rangle
= \sum_i U_{\alpha' i}\,~e^{-
iE_it}\,~U_{\alpha i}^*\,.
\label{017}
\end{equation}
where  $t$ is the transition time.

We will numerate neutrino masses in such a way that $$m_{1}
<m_{2}< m_{3}\,. $$ From (\ref{017}) we have for the probability
of the transitions
 $\nu_{\alpha}\to\nu_{\alpha '} $

\begin{equation}
P(\nu_\alpha\to\nu_{\alpha'})
=
\left|
\sum_{i=1}^{3} U_{{\alpha'} i} \,
 e^{ - i
 \,
\Delta{m}^2_{i1} \frac {L}{2 E} }
 U_{{\alpha}i}^*\,\right|^2 \,,
\label{018}
\end{equation}
where $\Delta{m}^{2}_{ik}=m^{2}_{i}-m^{2}_{k}$, $L\simeq t $ is
the distance between neutrino source and neutrino detector, $ E$
is the energy of the neutrinos.

Using the unitarity of the mixing matrix, we can rewrite this
expression in the following form

\begin{equation}
P(\nu_\alpha\to\nu_{\alpha'})
=
\left|
\delta_{\alpha \alpha'}
+
\sum_{i=2,3} U_{{\alpha'} i} \,
\left( e^{ - i
 \,
\Delta m^{2}_{i1} \frac {L}{2 E} } - 1 \right) U_{{\alpha}i}^*
\right|^2\,.
\label{019}
\end{equation}

We will assume that $ \Delta{m}^{2}_{21}$ is relevant for the
oscillations of the solar neutrinos and $ \Delta{m}^{2}_{31}$ is
relevant for the oscillations of atmospheric neutrinos.
\footnote{Another possibility,
the so-called inverted hierarchy, we will discuss later.} As we have
seen before, from the analysis of the existing solar and
atmospheric neutrino data follows that neutrino mass-squared
differences satisfy the hierarchy 
\begin{equation}
\Delta{m}^{2}_{21}\ll \Delta{m}^{2}_{31}.
\label{020}
\end{equation}

Let us first consider oscillations in the atmospheric and long
baseline (LBL) experiments. In these experiments $\frac{L}{E}\lesssim
10^{3}\,~\frac{ km}{\rm{GeV}}$ and the inequality
$$
\Delta m^{2}_{21} \frac {L}{2 E}\ll 1 $$

is satisfied.
Thus, the  contribution of the $i=2$ term to the expression
(\ref{019}) for the transition probability can be neglected. For
the  probability of the transition  $\nu_{\alpha}\to\nu_{\alpha '} $
($\alpha \neq \alpha'$)
we have (see, for example, \cite{BGG})

\begin{equation}
 P(\nu_{\alpha} \to \nu_{\alpha'}) =
\frac {1} {2} {\mathrm A}_{{\alpha'};\alpha}\,~ (1 - \cos\, \Delta m^{2}_{31} \frac
{L} {2E})\,,
\label{021}
\end{equation}

where the oscillation amplitude ${\mathrm A}_{{\alpha'};\alpha}$ is given by
\begin{equation}
{\mathrm A}_{\alpha'; \alpha} ={\mathrm A}_{\alpha; \alpha'}= 4 |U_{{\alpha'}3}|^2
|U_{{\alpha}3}|^2\,.
\label{022}
\end{equation}

The $\nu_{\alpha}$ survival probability can be obtained from
(\ref{021}) and the condition of the conservation of the
probability

$$ P(\nu_{\alpha} \to \nu_{\alpha}) = 1 - \sum _{\alpha'\neq \alpha }
 P(\nu_{\alpha} \to \nu_{\alpha'})\,.$$

We have

\begin{equation}
{\mathrm P}(\nu_\alpha \to \nu_\alpha) =
 1 - \frac {1} {2}{\mathrm B}_{\alpha ; \alpha}\,~ (1 - \cos \frac {\Delta m^2_{31} L} {2E})\,,
\label{023}
\end{equation}

where the oscillation amplitude ${\mathrm B}_{\alpha ; \alpha}$ is given by
\begin{equation}
{\mathrm B}_{\alpha ; \alpha}=
 \sum_{\alpha'\neq \alpha }
 {\mathrm A}_{{\alpha'};\alpha}=
 4 |U_{{\alpha}3}|^2\,~ ( 1 - |U_{{\alpha}3}|^2)
\label{024}
\end{equation}

It is obvious from (\ref{024}) that for the oscillation amplitudes
we have

$$ 0\leq{\mathrm B}_{\alpha ; \alpha}\leq 1;\,~
0\leq {\mathrm A}_{{\alpha'};\alpha}\leq 1$$.

Thus, due to the hierarchy relation (\ref{020}), the probabilities
of the transition $\nu_{\alpha} \to \nu_{\alpha'}$ in the
atmospheric and LBL experiments are determined by the largest
neutrino mass- squared difference $\Delta m^{2}_{31}$ and by the
elements $|U_{{\alpha}3}|^2$, which connect the flavor neutrino
fields $\nu_{\alpha L}$ with the field of the heaviest neutrino
$\nu_{3 L}$. Every oscillation channel is characterized by its own
oscillation amplitude. \footnote{Let us notice that in the case of
oscillations between two types of neutrinos from unitarity of the mixing matrixit follows that  oscillation
amplitudes are connected by the relations $ {\mathrm
A}_{{\alpha'};\alpha}= {\mathrm A}_{{\alpha};\alpha'}= {\mathrm
B}_{\alpha ; \alpha}= {\mathrm B}_{\alpha' ; \alpha'}=
 \sin^{2}  2\theta\,~ (\alpha'\neq \alpha ) $. }
Oscillation amplitudes in appearance and disappearance channels
are connected by the relation (\ref{024}).

From the unitarity of the mixing matrix follows that
$\sum_{\alpha}|U_{{\alpha}3}|^2 = 1$. Thus, in the leading
approximation, we are considering, all oscillation channels are characterized  by
three parameters. We can choose the parameters $$\Delta
m^{2}_{31}\,,~~ \tan^{2}\theta _{23}\,,~~  |U_{{\alpha}3}|^2$$

The oscillation amplitudes in the appearance and disappearance
channels are given by the expressions
\begin{equation}
{\mathrm A}_{\tau;\mu}= ( 1 - |U_{e3}|^2)^2\,\sin^{2}2\theta
_{23}\,;\,~~ {\mathrm A}_{e;\mu}=4\, |U_{e3}|^2 \,( 1 -
|U_{e3}|^2)\,\sin^{2}\theta _{23} 
\label{025}
\end{equation}
and
\begin{equation}
{\mathrm B}_{\mu;\mu}={\mathrm A}_{\tau;\mu}+{\mathrm A}_{e;\mu}\,~~~
{\mathrm B}_{e;e}=4\, |U_{e3}|^2 \,( 1 - |U_{e3}|^2)\,.
\label{026}
\end{equation}

Let us note that the following
relation exists between the oscillation amplitudes

$${\mathrm A}_{e;\mu}={\mathrm B}_{e;e}\,\sin^{2}\theta _{23}$$

The CP-violating phase  $\delta $ does not enter into the expressions
 (\ref{021}) and (\ref{022}) for the transition probabilities. Thus,
 in the leading approximation with the dominance of the $\Delta m^{2}_{31}$
term
the relations
$$ P(\nu_{\alpha} \to \nu_{\alpha'})=  P(\bar\nu_{\alpha} \to
\bar\nu_{\alpha'}) $$ are satisfied independently of the value of
the CP-violating phase. This means that the effect of CP violation
in the lepton sector can not be revealed if only one neutrino mass
squared difference is relevant for neutrino oscillations.

\section{Atmospheric neutrinos}

We will discuss now  the results of the atmospheric S-K experiment
\cite{AS-K}. In this experiment a significant up-down asymmetry of
the high-energy muon events was observed. If there are no neutrino
oscillations, the dependence of the number of the high energy muon
(electron) events on $\cos\,\theta_{z}$ must satisfy the  relation

\begin{equation}
N_{l}(\cos\theta_{z})= N_{l}( -\cos \theta_{z})\,~~ (l=e,\mu) \,,
\label{027}
\end{equation}
where $\theta_{z}$ is the zenith angle.

The measured dependence of the number of electron events on
$\cos\theta_{z}$ is in good agreement with (\ref{027}). For the
muon events in the Multi-GeV region ($E \geq 1.3\,~ \rm{GeV}$) a
strong $\cos\theta_{z}$ asymmetry was observed. For the ratio of
the total number $U$ of up-going muons ($\cos\theta_{z}\leq 0$) to
the total number $D$ of down-going muons ($\cos\theta_{z}\geq 0$)
was obtained

$$
\left(\frac{U}{D}\right)_{\mu} = 0.54 \pm 0.04 \pm 0.01\,.$$

The  $\cos\theta_{z}$ dependence of the number of muon events,
observed in the Super-Kamiokande experiment, is in a agreement
with the disappearance of muon neutrinos due to neutrino
oscillations. The data are perfectly described if the survival
probability has the two-neutrino form
\begin{equation}
 P(\nu_{\mu} \to \nu_{\mu}) = 1-
\frac {1} {2} \sin^{2}2\,\theta_{\rm{atm}}\,~ (1 - \cos\, \Delta m^{2}_{\rm{atm}} \frac
{L} {2E})
\label{028}
\end{equation}

From the analysis of the S-K data
the best-fit values (\ref{001}) of the neutrino oscillation parameters were obtained.

These values
are directly connected with the observed zenith angle dependence.
In fact, for the high-energy neutrinos the distance $L$ between
the region, where neutrinos are produced in the atmosphere, and
the detector is determined by the zenith angle $\theta_{z}$.
Down-going neutrinos with $\cos\theta_{z} = 1$ travel a distance
of about 20 km and up-going neutrinos with $\cos\theta_{z} =- 1$
travel a distance of about 13000 km. For the down-going neutrinos
the argument of the cosine in the expression (\ref{028}) for the
survival probability is small and
\begin{equation}
P^{down}(\nu_{\mu} \to \nu_{\mu})\simeq 1\,.
\label{029}
\end{equation}
For the up-going neutrinos the argument of the cosine in Eq.
(\ref{028}) is large and due to averaging over neutrino energies
and distances the cosine term in Eq. (\ref{028}) vanishes. For the
averaged survival probability we have
\begin{equation}
 P^{up}(\nu_{\mu} \to \nu_{\mu})\simeq
 1 -\frac{1}{2} \sin^{2}2\,\theta_{\rm{atm}}\,.
\label{030}
\end{equation}
The number of muon events with $\cos \theta_{z}\simeq 1$, observed in the S-K experiment, is about two times larger than the number of the muon events with
$\cos \theta_{z}\simeq -1$. Thus,
$P^{up}(\nu_{\mu} \to \nu_{\mu})\simeq 0.5$
and $ \sin^{2}2\,\theta_{\rm{atm}} \simeq 1$.

All data of the S-K experiment are in a good agreement with the
assumption of $|U_{e3}|^2=0$ and two-neutrino $\nu_{\mu} \to \nu_{\tau}$
oscillations. From the analysis of the data the following ranges
were obtained for the oscillation parameters
$$\sin^{2}2\,\theta_{\rm{atm}}\geq 0.88\,;~~ 1.6 \cdot 10^{-3}\leq
\Delta m^{2}_{\rm{atm}}\leq 4 \cdot 10^{-3}\rm{eV}^{2}\,.$$ 

The three-neutrino analysis of S-K atmospheric neutrino data
 allows
to obtain an upper bound for the parameter $|U_{e3}|^2$. In Ref.
\cite{Kaji} it was found that

$$|U_{e3}|^2 \leq 0.35\,.$$

\section{ The upper bound of the parameter $|U_{e3}|^2 $}

The most direct and stringent bound on the parameter $|U_{e3}|^2 $
can be obtained from the results of the LBL reactor experiments
CHOOZ \cite{CHOOZ} and Palo Verde \cite{PaloV} which are sensitive
to the atmospheric range of neutrino mass-squared differences. In
these experiments, $\bar \nu_{e}$'s from the reactors at a
distance of about 1 km from the detectors were recorded via the
observation of the process $$\bar \nu_{e} + p \to e^+ + n\,. $$.

No indications in favor of a disappearance of the reactor
$\bar\nu_{e}$'s were found. For the ratio $R$ of the total number
of the detected and expected events was obtained

$$ R = 1.01 \pm 2.8 \% (\rm{stat})\pm 2.7 \% (\rm{syst})\,~~~ \rm{CHOOZ}$$

and

$$ R = 1.01 \pm 2.4 \% (\rm{stat})\pm 5.3 \% (\rm{syst})\,~~~ \rm{Palo\,~ Verde}.$$

In the leading approximation (neglecting the contribution of the
$i=2$ term in Eq.(\ref{019})), we have for the $\bar \nu_{e}$
survival probability

\begin{equation}
{\mathrm P}(\bar\nu_e \to \bar\nu_e) =
 1 - \frac {1} {2}{\mathrm B}_{e ; e}\,~ (1 - \cos \frac {\Delta m^2_{31} L} {2E})\,,
\label{031}
\end{equation}

where
$$
{\mathrm B}_{e; e}=
 4 |U_{e3}|^2\,~ ( 1 - |U_{e3}|^2)\,.$$

From the exclusion plot, obtained from the analysis of the data of
the CHOOZ (or Palo Verde) experiment at a fixed value of $\Delta
m^2_{31}$ for the allowed values of the oscillation amplitude
 we have the bound
\begin{equation}
{\mathrm B}_{e ; e}\leq{\mathrm B}^{0} _{e ; e}\,.
 \label{032}
\end{equation}
 This bound
depends on the value of $\Delta m^2_{31}$. From (\ref{031}) and
(\ref{032}) follows

\begin{equation}
|U_{e 3}|^{2} \leq
\frac{1}{2}\,\left(1 - \sqrt{1-{\mathrm B}_{e;e}^{0} }\right)
\label{033}
\end{equation}

or
\begin{equation}
|U_{e 3}|^{2} \geq
\frac{1}{2}\,\left(1 + \sqrt{1-{\mathrm B}_{e;e}^{0} }\right)\,.
\label{034}
\end{equation}
We are interested in the region of $\Delta m^2_{31}\gtrsim \,
10^{-3}\rm{eV}^{2}$. In this region the amplitude ${\mathrm
B}_{e;e}^{0}$ is small. Thus, $|U_{e 3}|^{2}$ can be small
(inequality (\ref{033})) or  large, close to one (inequality
(\ref{034})). This last possibility is excluded by the solar
neutrino data. In fact, as we will see later, the heaviest neutrino $\nu_{3}$ (both in
the case of vacuum  and in the case of matter 
) gives an incoherent contribution  $|U_{e 3}|^{4}$ 
to the survival probability of the solar $\nu_{e}$.
If $ |U_{e 3}|^{2}$ is large, the probability of solar neutrinos
to survive will be close to one. As we will see in the next
section this possibility is excluded in a model independent way by
S-K and SNO data.

Thus, taking into account solar neutrino data, we can conclude
from the results of the LBL reactor experiments that the parameter
$ |U_{e 3}|^{2} $ is small. From the CHOOZ exclusion curve at
$\Delta m^2_{31}= 2.5\cdot 10^{-3}\rm{eV}$ (the S-K best-fit
value) we have

\begin{equation}
|U_{e 3}|^{2}\leq 3.7 \cdot 10^{-2}\,. \label{035}
\end{equation}

This bound was obtained under the assumption that the contribution
of the $i=2$ term to the expression for the transition probability
in Eq.(\ref{019}) can be neglected in the LBL region. This is a
good approximation for the LMA best-fit value of
 $\Delta m^{2}_{\rm{sol}}$
given by Eq.\,(\ref{002}). However, the values of $\Delta
m^{2}_{\rm{sol}}$ in the LMA region (which is the preferable fit
to the solar neutrino data) can be as large as $\Delta
m^{2}_{21}\simeq 6\cdot 10^{-4}\rm{eV}^{2}  $ (see, for example,
\cite{Bahcall}). In the CHOOZ experiment the average value of the
parameter $\frac{L}{ E}$ is approximately equal to 300 m/MeV. If,
for example, At $\Delta m_{\rm{sol}}^{2}= 2\cdot 10^{-4}\rm{eV}^{2}$, 
for example,
for the average value of the parameter $\frac{L}{ E}$ in the
CHOOZ experiment we have

$$\Delta m_{\rm{sol}}^{2}\,\frac {L} {2E}  \simeq 1.5\cdot 10^{-1}\,. $$

Thus, the i=2 term in Eq.(\ref{019}) at the relatively large
values of $\Delta m_{21}^{2}$, belonging to the LMA region, could
give a sizable contribution to the survival probability.

From Eq.~(\ref{019}), taking into account all terms,
we obtain the following
expression for the $\bar \nu_{e}$
survival probability \cite{BNP}

\begin{eqnarray}
\lefteqn{P({\bar \nu_e}\to{\bar \nu_e})} \nonumber\\
&& =\,~~ 1 - 2 \, |U_{e 3}|^2 \left( 1 - |U_{e 3}|^2 \right)
\left( 1 - \cos \frac{ \Delta{m}^2_{31} \, L }{ 2 \, E } \right)
\nonumber \\
&& -\,~~\frac{1}{2} \,(1- |U_{e 3}|^2)^{2}\sin^{2}2\,\theta _{\rm{sol}}
\,
\left( 1 - \cos \frac{ \Delta{m}^2_{\rm{sol}} \, L }{ 2 \, E } \right)   
\label{036}\\
& & +\,~~ 2\, |U_{e 3}|^2 \,(1- |U_{e 3}|^2)\,\sin^{2}\theta _{\rm{sol}}\,  
\left(\cos
\left( \frac
{\Delta{m}^2_{31} \, L }{ 2 \, E} - \frac {\Delta{m}^2_{\rm{sol}} \, L }{ 2 \,
E}\right)
-\cos \frac {\Delta{m}^2_{31} \, L }{ 2 \, E} \right)\,, \nonumber
\end{eqnarray}

where we put $\Delta{m}^{2}_{21}=\Delta{m}^{2}_{\rm{sol}}$ and
$\theta _{12}=\theta _{\rm{sol}}$.

The second term in the right-hand part of this expression comes
from the main i = 3 term in Eq.(\ref{019}), the third one from
the i = 2 term and the fourth one from the interference of the i =3
and i = 2 terms.

Let us note that in the LMA region the angle $\theta _{\rm{sol}}$
is large. Thus, the coefficient in front of the bracket of the
third ``solar term'' in Eq.(\ref{036}) is large  and the
coefficient in front of the bracket of the interference term is of
the same order as the coefficient in front of the bracket of the
second term.

Up to now we have assumed that the mass-squared difference
$\Delta{m}^{2}_{21}$ of the lightest neutrinos $\nu_{2}$ and
$\nu_{1}$ is relevant for the oscillations of the solar neutrinos
and hierarchy (\ref{020}) is valid. The existing data do not
exclude, however, the possibility that the mass-squared difference
$\Delta{m}^{2}_{32}$ between the squares of the masses of the
heaviest neutrinos $\nu_{3}$ and  $\nu_{2}$
 is relevant for
the oscillations of the solar neutrinos and $\Delta{m}^{2}_{31}$
is relevant for the oscillations of the atmospheric neutrinos. In
this case the so-called inverted hierarchy
\begin{equation}
\Delta{m}^{2}_{32}\ll \Delta{m}^{2}_{31}
\label{037}
\end{equation}
takes place.\footnote{Recall that we numerate neutrino masses in
such a way that $m_{1} <m_{2}< m_{3} $.}

We will consider now the case of the inverted neutrino mass spectrum.
The probability of the transition
$\nu_\alpha\to\nu_{\alpha'}$ in vacuum can be written in the form
\begin{equation}
P(\nu_\alpha\to\nu_{\alpha'})
=
\left|
\delta_{\alpha \alpha'}
+
\sum_{i=1,2} U_{{\alpha'} i} \,
\left( e^{ i
 \,
\Delta m^{2}_{3i} \frac {L}{2 E} } - 1 \right) U_{{\alpha}i}^*
\right|^2\,.
\label{038}
\end{equation}

Let us consider  neutrino oscillations in the atmospheric and LBL
experiments. Neglecting the contribution of the small $i=2$ term
in the expression (\ref{038}), for the transition probabilities we
will obtain expressions (\ref{021}) and (\ref{023}). The
oscillation amplitudes ${\mathrm A}_{{\alpha'};\alpha}$ and
${\mathrm B}_{\alpha ; \alpha}$ are given by the expressions
(\ref{022}) and (\ref{024}) after applying the change $|U_{\alpha
3}|^2 \to |U_{\alpha 1}|^2$. Thus, in the case of the inverted
hierarchy, in the leading approximation, the transition
probabilities in the atmospheric and LBL experiments are
determined by the largest neutrino mass-squared difference
$\Delta{m}^2_{31}$ and the elements $|U_{\alpha 1}|^2$, connecting
the field of the flavor neutrino $\nu_{\alpha L}$ with the field
of the lightest neutrino $\nu_{1L}$. In order to obtain the
three-neutrino expression for the $\bar \nu_{e}$ survival
probability in the case of the inverted hierarchy it is necessary
to change $|U_{e3}|^2 \to |U_{e 1}|^2$ and $\theta_{\rm{sol}} \to
\pi/2 - \theta_{\rm{sol}}$ in Eq.(\ref{036}).

The  expression (\ref{036}) and similarly the expression for the
$\bar \nu_{e}$ survival probability in the case of the inverted
hierarchy were used in Ref. \cite{BNP} in order to obtain from the
CHOOZ data new exclusion plots in the plane of the parameters
$|U_{e 3}|^2$ - $\Delta{m}^{2}_{31}$ (and $|U_{e 1}|^2$ -
$\Delta{m}^{2}_{31}$ in the case of inverted hierarchy) at fixed
values of $\Delta{m}^2_{\rm{sol}}$ and $\sin^{2}\,\theta
_{\rm{sol}}$, belonging to the LMA allowed region.

In Table 1 we present the upper bounds of the parameter $|U_{e
3}|^2$ ($|U_{e 1}|^2$) at $\Delta{m}^{2}_{31}= 2.5 \cdot
10^{-3}\rm{eV}^{2}$ and $\Delta{m}^{2}_{31}= 10^{-2}\rm{eV}^{2}$.
As can be seen from Table 1, at $\Delta{m}^{2}_{\rm{sol}}\leq
2\cdot 10^{-4}\rm{eV}^{2}$ the bounds on $|U_{e 3}|^2$ ($|U_{e
1}|^2$) in the  three-neutrino case are practically the same as in
the leading approximation which corresponds to the two-neutrino
case. At the larger values of $\Delta{m}^{2}_{\rm{sol}}$ the
limits on $|U_{e 3}|^2$ ($|U_{e 1}|^2$) are more stringent in the
three-neutrino case than in the two-neutrino case. For example, at
$\Delta{m}^{2}_{\rm{sol}}= 6\cdot 10^{-4}\rm{eV}^{2}$ and
$\sin^{2}\,\theta _{\rm{sol}}= 0.27 $ at the S-K best-fit point
$\Delta{m}^{2}_{31}= 2.5 \cdot 10^{-3}\rm{eV}^{2}$ we have
\begin{equation}
|U_{e 3}|^2\leq 2\cdot 10^{-2}\,.
\label{039}
\end{equation}

This bound is about 2 times smaller that the
upper bound (\ref{035}), obtained
in the two-neutrino case.

New data of the solar neutrino exeriments (SNO, GNO, BOREXINO
\cite{ BOR}) and data of the reactor experiment KamLAND
\cite{KamLAND} will allow to determine the values of the
parameters $\Delta{m}^{2}_{\rm{sol}}$ and $\tan^{2}\,\theta
_{\rm{sol}}$ with better accuracy than today. That will permit to
obtain from the results of the CHOOZ and Palo Verde experiments
more precise upper bounds of the parameter $|U_{e 3}|^2$. Let us
stress that the exact value of the parameter $|U_{e 3}|^2$ is very
important for the future Super Beam (see \cite{jhf}) and Neutrino
Factory (see \cite{Cline}) experiments, in which
$\nu_{\mu}\to\nu_{e}$ oscillations and effects of the
three-neutrino mixing, in particular effects of CP violation in
the lepton sector, will be investigated in detail.

\begin{table}[htbp]
  \caption{\small Upper bounds of the mixing parameter $|U_{e3}|^2$
($|U_{e1}|^2$ in the case of inverted hierarchy)
 for different values of  neutrino
oscillation parameters.}
  \label{tab:mixlim}
  \begin{center}
    \begin{tabular}{|c|c|c|c|c|}
      \hline
        $\Delta m^2_{31}$ & $\Delta m^2_{\rm{sol}}$ &
        $|U_{e3}|^2$ & $|U_{e3}|^2$ & $|U_{e1}|^2$ \\
        $({\rm eV}^2)$ & $({\rm eV}^2)$ &
        $(\sin^2 \theta_{\rm{sol}} = 0.5)$ & $(\sin^2 \theta_{\rm{sol}}
         = 0.27)$ &
        $(\sin^2 \theta_{\rm{sol}} = 0.27)$ \\
      \hline
      \hline
        $2.5\cdot 10^{-3}$ & $0$ &
        \multicolumn{3}{c}{$3.6\cdot 10^{-2}$} \vline \\
      \cline{2-5}
        & $2\cdot 10^{-4}$ & $3.6\cdot 10^{-2}$ & $3.5\cdot 10^{-2}$
        & $3.7\cdot 10^{-2}$ \\
      \cline{2-5}
        & $4\cdot 10^{-4}$ & $2.9\cdot 10^{-2}$ & $3.0\cdot 10^{-2}$
        & $3.4\cdot 10^{-2}$ \\
      \cline{2-5}
        & $6\cdot 10^{-4}$ & $1.7\cdot 10^{-2}$ & $2.0\cdot 10^{-2}$
        & $2.5\cdot 10^{-2}$ \\
      \hline
      \hline
        $10^{-2}$ & $0$ &
        \multicolumn{3}{c}{$3.6\cdot 10^{-2}$} \vline \\
      \cline{2-5}
        & $2\cdot 10^{-4}$ & $3.4\cdot 10^{-2}$ & $3.4\cdot 10^{-2}$
        & $3.4\cdot 10^{-2}$ \\
      \cline{2-5}
        & $4\cdot 10^{-4}$ & $2.7\cdot 10^{-2}$ & $2.9\cdot 10^{-2}$
        & $2.8\cdot 10^{-2}$ \\
      \cline{2-5}
        & $6\cdot 10^{-4}$ & $1.7\cdot 10^{-2}$ & $2.1\cdot 10^{-2}$
        & $2.0\cdot 10^{-2}$ \\
      \hline
      \hline
    \end{tabular}
  \end{center}
\end{table}

\section{Solar neutrinos}

We now come to the discussion of solar neutrinos. The energy of
the sun is generated in the reactions of the thermonuclear pp and
CNO cycles. From a thermodynamical point of view the energy of the
sun is produced in the transition of four protons and two
electrons into $^4$He and two electron neutrinos
\begin{equation}
4 \, p + 2 \, e^-
\to
{}^4He + 2 \, \nu_e
\,.
\label{040}
\end{equation}
The energy, which is released in this transition,
is equal to
$$
Q = 4 m_p + 2 m_e - m_{^4He} \simeq 26.7 MeV\,.
$$

Thus, the production of the energy of the sun is {\em accompanied}
by the emission of $\nu_e$'s.

From (\ref{040}) we can obtain a model independent
relation that connects the fluxes of neutrinos with
the luminosity of the sun $L_\odot$.
In fact, from (\ref{040}) it follows
that the luminous energy per one neutrino with energy $E$ is equal to
$\frac{1}{2}\,~(Q-2E)$.
Thus, the total luminosity of the sun is given by

\begin{equation}
\int {\frac{1}{2} (Q -2 E)\,~N(E)dE} =
L_\odot \,,
\label{041}
\end{equation}
where  $N(E)$ is
the total number of neutrinos with the energy $E$,
 produced by the sun in 1 sec.

We have
\begin{equation}
N(E) = 4 \pi R^{2} \sum_i \Phi_{i}(E)
\label{042}
\end{equation}
where $ R$
is the distance between the sun and the earth and
$\Phi_{i}(E)$ is the flux
of $\nu_{e}$ from the source $i$
(in the case of neutrino oscillations $\Phi_{i}(E)$ is
the total flux of all types of neutrinos, including sterile neutrinos).

From (\ref{041}) and (\ref{042}) we obtain the
luminosity relation
\begin{equation}
\sum_{i}(\frac{1}{2}Q -\overline E_{i})\Phi_{i}^{0}
=\frac{L_\odot} {4 \pi R^2}\,,
\label{043}
\end{equation}
where $\Phi_{i}^{0}$ is the total flux and  $\overline E_{i}$
is the average energy of neutrinos
from the source
$i$.

Let us note that in the derivation of the luminosity relation (\ref{043})
we
assumed that the sun is in a stable state. \footnote{ It takes
about $7\cdot 10^{5}$ years for the photons, produced in the
central zone of the sun, to reach the surface (see \cite{JB}).}

The main source of solar neutrinos is the $pp$ reaction
\begin{equation}
p+p \to d + e^- + \nu_{e}\,.
\label{044}
\end{equation}
This reaction is the source of the low energy neutrinos with a
maximum energy of  0.42 MeV. The total flux of  $pp$ neutrinos,
predicted by the Standard Solar Model
 BP00 (SSM BP00)
\cite{BPin},
 is determined mainly by the luminosity relation (\ref{043})
and is equal to
$\Phi_{pp}= 5.95 \cdot 10^{10}cm^{-2}s^{-1}$.

The reaction
\begin{equation}
e^{-} + ^7\rm{Be} \to ^7\rm{Li} +\nu_{e}\,.
\label{045}
\end{equation}
is the source of the  monochromatic neutrinos with an energy of
0.86 MeV. The flux of $^7$Be neutrinos, predicted by the SSM, is
$\Phi_{^7 Be}= 4.8 \cdot 10^{9}cm^{-2}s^{-1}$

In the Super-Kamiokande and the SNO experiments,
mainly high energy neutrinos from the decay
\begin{equation}
^8\rm{B} \to ^8\rm{Be}^{*} +e^{+}+\nu_{e}\,.
\label{046}
\end{equation}
are detected (high energy thresholds). The maximum energy of the
$^8$B neutrinos is approximately equal to 15 MeV and the flux,
predicted by the SSM BP00, is $\Phi_{^8 B}= 5.9 \cdot
10^{6}cm^{-2}s^{-1}$.

At present results of six solar neutrino experiments (Homestake
\cite{Cl}, GALL-EX-GNO \cite{GALLEX,GNO}, Kamiokande \cite{Kam} ,
Super-Kamiokande \cite{S-K} and SNO\cite{SNO}) are available. The
event rates, measured in all solar neutrino experiments, are
significantly smaller that the rates predicted by the SSM.

In the Homestake experiment solar neutrinos are detected by the
radiochemical method through the observation of the reaction

$$\nu_{e}+ ^{37}\rm{Cl}\to e^{-}+^{37}\rm{Ar} \,. $$

The threshold of this process is $E_{th}= 0.81 \rm{MeV}$. Thus, in
this experiments mainly $^8\rm{B}$ and $^7\rm{Be}$ neutrinos are
detected. The observed event rate is equal to \footnote{1\,~SNU =
$10^{-36}$events\,~ atom $^{-1}$\,~ s$^{-1}$.}

$$ R_{Cl} = (2.56 \pm 0.16 \pm 0.16)\,~ SNU $$

The rate, predicted by SSM BP00,
is

$$ R_{Cl}^{SSM} =(8.59 \begin{array}{c} +1.1 \\-1.2\end{array})\,~
SNU$$

In the GALLEX-GNO  and SAGE experiments solar neutrinos are
detected through the observation of the reaction $$\nu_{e}+
^{71}\rm{Ga}\to e^{-}+^{71}\rm{Ge}\,.  $$ The threshold of this
reaction is $E_{th}= 0.23\,~ \rm{MeV}$. Thus, in the Gallium
experiments neutrinos from all reactions in the sun are detected.
The combined event rate of the GNO and the GALLEX experiment is

$$ R_{Ga} = (74.1 \pm 6.7 \pm 6.8)\,~ SNU\,;\,~~~ GALLEX-GNO $$
The event rate measured in the SAGE experiment is $$ R_{Ga} = (77
\pm 6 \pm 3)\,~ SNU\,;\,~~~ SAGE $$

The rate, predicted by SSM BP00 for the Gallium experiments, is

$$ R_{Ga}^{SSM} =(130 \begin{array}{c} +9 \\-7 \end{array})\,~
SNU\,.$$

\section{Comparison of the results of the SNO and the S-K experiments}

The Kamiokande, Super-Kamiokande and SNO are direct counting
experiments. In the Kamiokande and S-K experiments the solar
neutrinos are observed through the detection of the recoil
electrons in the elastic (ES) neutrino-electron
 scattering
\begin{equation}
\nu + e \to \nu + e \,.
\label{047}
\end{equation}
We will discuss the results of the  S-K experiment \cite{S-K}. In
this experiment the large 50 kton water Cherenkov detector is
used. During 1258 days of running, in the S-K experiment $18464
\begin{array}{c} +677 \\-590\end{array}$ events with the energy of
the recoil electrons in the range 5-20 $\rm{MeV}$ were observed.

As the energy of the recoil electrons is much larger than the
electron mass, the direction of the recoil electron  momentum
practically coincides with the direction of the neutrino momentum.
In the S-K experiment, in the distribution of the events on
$\cos\theta_{sun}$ ($\theta_{sun}$ is the angle between the recoil
electron momenta and the direction to the sun) a sharp peak at
$\cos\theta_{sun}=1$ was observed. The observation of such a peak
is a clear demonstration that the recorded events are due to solar
neutrinos.

All flavor neutrinos
$\nu_e$, $\nu_\mu$ and $\nu_\tau$ are detected in the S-K experiment. However,
the  cross
section
of the (NC) $\nu_{\mu}$ ($\nu_{\tau}$)  - $e$ scattering is about six times smaller
than the cross section of the (CC+NC) $\nu_e-e$ scattering.
Thus, the S-K sensitivity to $\nu_{\mu}$ and $\nu_{\tau}$ is much lower than
the sensitivity to $\nu_{e}$.

In the SNO experiment \cite{SNO}
a heavy water Cherenkov detector is used (1 kton of
$\rm{D}_{2}\rm{O}$). Solar neutrinos were detected in the experiment via the
observation
of the CC reaction
\begin{equation}
\nu_e + d \to e^{-}+ p +p
\label{048}
\end{equation}
and the elastic scattering reaction
\begin{equation}
\nu + e \to \nu + e\,.
\label{049}
\end{equation}

The electron kinetic energy threshold in the SNO experiment is
$T_{th}$=6.75 MeV. Thus, in the SNO experiment, like in the S-K
experiment, only high energy $^{8}\rm{B}$ neutrinos are
detected.\footnote{ High energy neutrinos from the hep reaction
($p+^3\rm{He}\to ^4\rm{He}+ e^{+}+\nu_e$) give a very small
contribution to the event rate.} From November 1999 till January
2001  $975.4 \pm 39.7$ CC events and $106.1 \pm 15.2$ ES events
were observed.

The measured ES event rate is consistent with the more precise S-K
event rate. The detection of the solar neutrinos via the
observation of the CC reaction (\ref{048}) allowed to determine
the flux of the solar $\nu_{e}$ on the earth. In fact, the total
event rate, measured in the SNO experiment, via the observation of
the CC reaction is given by
\begin{equation}
R^{CC} = \int_{E_{0}}\,~ \sigma_{\nu_{e}d}(E)\,~\Phi_{\nu_{e}}(E)\,d\,E \,,
\label{050}
\end{equation}
where  $\sigma_{\nu_{e}d}(E)$ is the total cross section of the CC
process \footnote{This quantity depends on $T_{th}$
and on the resolution of the detector.}, $\Phi_{\nu_{e}}(E)$ is
the flux of the $\nu_{e}$ on the earth and $E_{0} = T_{th}+ 1.44 \rm{MeV}$.

The flux of $\nu_{e}$ on the earth is given by

\begin{equation}
\Phi_{\nu_{e}}(E) = P(\nu_{e}\to\nu_{e})\,~\Phi_{\nu_{e}}^{0}(E)\,,
\label{051}
\end{equation}
where $ P(\nu_{e}\to\nu_{e})$ is the probability of the solar
$\nu_{e}$ to survive and $\Phi_{\nu_{e}}^{0}(E)$ is the initial
flux of $\nu_{e}$ (the flux that would be observed if there would
be no neutrino oscillations).

The flux $\Phi_{\nu_{e}}^{0}(E)$ can be presented in the form

\begin{equation}
\Phi_{\nu_{e}}^{0}(E)= X(E)\,~\Phi_{\nu_{e}}^{0}\,,
\label{052}
\end{equation}
where $\Phi_{\nu_{e}}^{0}$ is the total initial flux of the
$^{8}\rm{B}$ neutrinos and $X(E)$ is a normalized function ($\int
X(E)\,d\,E =1 $). The function $X(E)$ characterizes the spectrum
of $\nu_{e}$ from the decay $^8\rm{B} \to ^8\rm{Be}^{*}
+e^{+}+\nu_{e}$. This function is known.

Let us determine the function $\rho_{\nu_{e}d} (E)$ by the relation
\begin{equation}
\sigma_{\nu_{e}d}(E)\,~X(E) =<\sigma_{\nu_{e}d}>\rho _{\nu_{e}d} (E)\,,
\label{053}
\end{equation}
where 
$$<\sigma_{\nu_{e}d}> =  \int_{E_{0}}\,~
\sigma_{\nu_{e}d}(E)\,~X(E)\,d\,E $$ 
is the averaged cross section of the CC process (\ref{048}).
It is obvious that
$$
\int_{E_{0}}\,~\rho _{\nu_{e}d}\,~d\,E =1\,.$$

From Eq.(\ref{050}) and Eq.(\ref{053}) for the  CC event rate we
have

\begin{equation}
R^{CC}=  <\sigma_{\nu_{e}d}>\Phi_{\nu_{e}}^{CC}\,,
\label{054}
\end{equation}

where
the average flux of $\nu_{e}$ on the earth $\Phi_{\nu_{e}}^{CC}$ is given by
\begin{equation}
\Phi_{\nu_{e}}^{CC} =
< P(\nu_{e}\to\nu_{e})>_{\nu_{e}d}\,~\Phi_{\nu_{e}}^{0}
\label{055}
\end{equation}
In this equation

\begin{equation}
< P(\nu_{e}\to\nu_{e})>_{\nu_{e}d} =  \int_{E_{0}}\,~ P(\nu_{e}\to\nu_{e})\rho _{\nu_{e}d}\,~d \,E
\label{056}
\end{equation}
is the averaged survival probability.

In the SNO experiment the spectrum of the produced electrons was
measured. No significant deviations from the predicted spectrum
were found. Thus the results of the SNO experiment are compatible
with the assumption that the $\nu_{e}$ survival probability in the
SNO energy region is constant. We have in this case
\begin{equation}
< P(\nu_{e}\to\nu_{e})>_{\nu_{e}d}\simeq  P(\nu_{e}\to\nu_{e})\,. 
\label{057}
\end{equation}

From the data of the SNO experiment for the averaged flux of
$\nu_{e}$ on the earth the following value was found \cite{SNO}

\begin{equation}
(\Phi_{\nu_{e}}^{CC})_{SNO} = (1.75 \pm 0.07 \pm 0.12 \pm 0.05\,~ (\rm{theor}))
 \cdot 10^{6}\,~
cm^{-2}s^{-1} \,.
\label{058}
\end{equation}

Let us now discuss the results of the S-K experiment. In this experiment
solar neutrinos are detected via the observation of the ES process
(\ref{047}). The total S-K event rate is given by
\begin{equation}
R^{ES}= \int_{E_{0}}\,~ \sigma_{\nu_{e}e}(E)\,~\Phi_{\nu_{e}}(E)\,d\,E
+  \int_{E_{0}}\,~ \sigma_{\nu_{\mu} e}(E)\,~\sum_{l=\mu,\tau} \Phi_{\nu_{l}}(E)\,d\,E\,,
\label{059}
\end{equation}
where $\sigma_{\nu_{l}e}(E)$ is the cross section of $\nu_{l}-e$
scattering, $\Phi_{\nu_{l}}(E)$ is the flux of $\nu_{l}$ on the
earth ($l = e,\mu,\tau$ ) and $E_{0}$ is given by 

\begin{equation}
E_{0}= \frac{T_{th}}{2}\,~ \left( 1 + \sqrt{1 +\frac{2m}{T_{th}}}\right).
\label{060}
\end{equation}
The total energy threshold in the S-K experiment is $E_{th} = 5
\rm{MeV}$.

For the flux of  $\nu_{l}$ on the earth we have
\begin{equation}
\Phi_{\nu_{l}}(E)= P(\nu_{e}\to\nu_{l})\,~\Phi_{\nu_{e}}^{0}(E)\,,
\label{061}
\end{equation}
where $\Phi_{\nu_{e}}^{0}(E)$ is the initial flux of $\nu_{e}$. This flux
is given by  Eq.(\ref{052}).

Let us determine the normalized functions $\rho _{\nu_{l}e} (E)$
($l= e, \mu $) as follows:

\begin{equation}
\sigma_{\nu_{l}e}(E)\,~X(E) =<\sigma_{\nu_{l}e}>\rho _{\nu_{l}e} (E)\,,
\label{062}
\end{equation}
where
\begin{equation}
\int_{E_{0}}\sigma_{\nu_{l}e}(E)\,~X(E)\,~d\,E
=<\sigma_{\nu_{l}e}>\,,\label{063}
\end{equation}
and
$$ \int_{E_{0}}\,~\rho _{\nu_{l}e}\,~d\,E =1.$$

From (\ref{060})-(\ref{062}) it follows that the total S-K event rate
can be written in the form
\begin{equation}
R^{ES}=  <\sigma_{\nu_{e}e}>\Phi_{\nu}^{ES}\,,
\label{064}
\end{equation}
where
\begin{equation}
\Phi_{\nu}^{ES}= \Phi_{\nu_{e}}^{ES} + \frac{<\sigma_{\nu_{\mu}e}>}
{<\sigma_{\nu_{e}e}>}\sum _{l=\mu,\tau }\Phi_{\nu_{l}}^{ES}\,.
\label{065}
\end{equation}
Here
\begin{equation}
\frac{<\sigma_{\nu_{\mu}e}>}{<\sigma_{\nu_{e}e}>} \simeq 0.154
\label{066}
\end{equation}
and the averaged fluxes of  $\nu_{e}$,  $\nu_{\mu}$ and $\nu_{\tau}$
 on the earth
are given by

\begin{equation}
\Phi_{\nu_{e}}^{ES} = <
P(\nu_{e}\to\nu_{e})>_{\nu_{e}e}\,\Phi_{\nu_{e}}^{0};\,~ \sum
_{l=\mu,\tau }\Phi_{\nu_{l}}^{ES} = \sum _{l=\mu,\tau }<
P(\nu_{e}\to\nu_{l})>_{\nu_{\mu}e}\,\Phi_{\nu_{e}}^{0}\,.
\label{067}
\end{equation}

From the data of the S-K experiment for the effective flux
$\Phi_{\nu}^{ES}$ it was obtained the value \cite{S-K} 

\begin{equation}
(\Phi_{\nu}^{ES})_{SK}= (2.32 \pm 0.03 \pm 0.08)
\cdot 10^{6}cm^{-2}s^{-1} \,.
\label{068}
\end{equation}

In the S-K experiment the spectrum of the recoil electrons was
measured. No deviation from the predicted spectrum was observed.
Thus, both the S-K and the SNO data are compatible with the
assumption of a constant $\nu_{e}\to\nu_{e}$ survival probability.
We have in this case

\begin{equation}
< P(\nu_{e}\to\nu_{e})>_{\nu_{e}e}\simeq  P(\nu_{e}\to\nu_{e})\,.
\label{069}
\end{equation}

Using (\ref{055}), (\ref{057}), (\ref{067}), and (\ref{069}) we
conclude that \footnote{It was shown in \cite{Villan}
 that it is possible to choose
the S-K and the SNO thresholds in such a way that these quantities
will be practically equal at any $ P (\nu_{e}\to\nu_{e})$. }

\begin{equation}
\Phi_{\nu_{e}}^{SNO}\simeq \Phi_{\nu_{e}}^{ES}
\label{070}
\end{equation}

From the comparison of the fluxes, measured in the S-K and SNO
experiments we can determine the flux of $\nu_{\mu}$ and
$\nu_{\tau}$ on the earth. From (\ref{058}), (\ref{065}),
(\ref{068}) and (\ref{070}), we have \cite{SNO}
\begin{equation}
\sum _{l=\mu,\tau }\Phi_{\nu_{l}}^{ES}= (3.69 \pm 1.13)\cdot 10^{6}\,~cm^{-2}
\,s^{-1}
\label{071}
\end{equation}
Thus, the results of the SNO and the S-K experiments give us the
first model independent evidence (at $ 3\sigma $ level) of the
presence of $\nu_{\mu}$ and $\nu_{\tau}$ in the flux of solar
neutrinos on the earth. The flux of $\nu_{\mu}$ and $\nu_{\tau}$
on the earth is approximately two times larger than the flux of
$\nu_{e}$.

For the total flux of the flavor neutrinos on the earth we obtain
from (\ref{058}), (\ref{070}) and (\ref{071})

\begin{equation}
\sum _{l=e, \mu,\tau }\Phi_{\nu_{l}}^{ES}=
(5.44 \pm 0.99)\cdot 10^{6}\,~cm^{-2}
\,s^{-1}\,.
\label{072}
\end{equation}

The SSM BP00 value of the total flux of $^8$B neutrinos is
\cite{Bahcall}

$$(\Phi_{\nu_{e}}^{0})_{SSM} = (5.93 \pm 0.89)\cdot 10^{6}\,~cm^{-2}.$$

Thus, the total flux of the flavor neutrinos, obtained from the
results of the SNO and S-K experiments, is in a good agreement
with the value of the total flux of $\nu_{e}$ predicted by the SSM
BP00.

Let us note that in the region of energies we are interested in we
have

$$\rho _{\nu_{\mu}e} (E) \simeq \rho _{\nu_{e}e} (E).$$

Taking this relation into account we find for the total flux of
flavor neutrinos on the earth

\begin{equation}
\sum _{l=e, \mu,\tau }\Phi_{\nu_{l}}^{ES}=\sum _{l=e, \mu,\tau }
< P(\nu_{e}\to\nu_{l})>_{\nu_{e}e}\,\Phi_{\nu_{e}}^{0}\,,
\label{073}
\end{equation}
where $\Phi_{\nu_{e}}^{0}$ is the total initial flux of $\nu_{e}$.
Thus, if there are no transitions of the solar  $\nu_{e}$ into sterile states,
we have
\begin{equation}
\sum _{l=e, \mu,\tau }\Phi_{\nu_{l}}^{ES}=\Phi_{\nu_{e}}^{0}\,.
\label{074}
\end{equation}

\section{Solar neutrino oscillations in the framework of three-neutrino mixing}

The data of all solar neutrino experiments are well described by
the two- neutrino transition probability, which is characterized
by the two parameters $\Delta m^{2}_{\rm{sol}}$ and
$\tan^{2}\theta_{\rm{sol}}$. We will now discuss the oscillations
observed in the solar neutrino experiments in the framework of
three neutrino mixing.

The probability of the transition $ \nu_{\alpha} \to \nu_{\alpha'}$
in vacuum (see (Eq.\ref{018})) can be presented in the
form

\begin{equation}
P(\nu_\alpha\to\nu_{\alpha'})
=
\left|
\sum_{i=1, 2} U_{{\alpha'} i} \,
 e^{ - i
 \,
\Delta{m}^2_{i1} \frac {L}{2 E} }
 U_{{\alpha}i}^* +  U_{{\alpha'} 3} \,
 e^{ - i
 \,
\Delta{m}^2_{31} \frac {L}{2 E} }
 U_{{\alpha}3}^*\,\right|^2 \,.
\label{075}
\end{equation}

We are interested in the probability, averaged over the region
where neutrinos are produced, over the neutrino spectrum, etc.
Because of the hierarchy (\ref{020}), in the expression for the
averaged transition probability
 the interference term does not enter.
 For the averaged transition probability we have

\begin{equation}
\overline P(\nu_\alpha\to\nu_{\alpha'})
=
\left|
\sum_{i=1, 2} U_{{\alpha'} i} \,
 e^{ - i
 \,
\Delta{m}^2_{i1} \frac {L}{2 E} }
 U_{{\alpha}i}^*\,\right|^2 + \left| U_{{\alpha'} 3}\,\right|^2 \,
\left| U_{{\alpha}3}\,\right|^2 \,.
\label{076}
\end{equation}

Using the unitarity of the mixing matrix, it is easy to show that
the averaged transition probabilities satisfy the relation

\begin{equation}
\sum_{\alpha'=e,\mu,\tau}\overline P(\nu_\alpha\to\nu_{\alpha'})=1.
\label{077}
\end{equation}

Using Eq.(\ref{010}),
the averaged probability of solar $\nu_{e}$ to survive
can be presented
in the form

\begin{equation}
\overline P(\nu_{e}\to\nu_{e})=|U_{{e} 3}|^{4}+ (1-|U_{{e} 3}|^{2})^{2}\,~
P^{(1,2)}(\nu_{e}\to\nu_{e})\,,
\label{078}
\end{equation}

where
\begin{equation}
P^{(1,2)}(\nu_{e}\to\nu_{e}) = 1 - \frac {1} {2} \sin^{2}2\,\theta_{12}\,~ (1 - \cos\, \Delta m^{2}_{21} \frac
{L} {2E})\,.
\label{079}
\end{equation}

The probability $P^{(1,2)}(\nu_{e}\to\nu_{e})$
has the same form as
the two-neutrino survival probability and
depends on the parameters $\Delta
m^{2}_{21}$ and $\sin^{2}2\,\theta_{12}$.
Let us stress, however,
that in the case of three-neutrino mixing, solar $\nu_{e}$
transfer into $\nu_{\mu}$ and $\nu_{\tau}$.

The expression  Eq.(\ref{078}) is valid also in the case
of matter \cite{Schramm}. In this case
$P^{(1,2)}(\nu_{e}\to\nu_{e})$ is the two-neutrino survival
probability in matter. In the calculation of this quantity the
density of the electrons $\rho_{e}(x)$ in the effective
Hamiltonian of the interaction of neutrino with matter must be
changed to $(1-|U_{e 3}|^{2})\,~\rho_{e}(x)$.

As we have seen before, from the data of the long baseline reactor
experiments CHOOZ and Palo Verde follows that the element $|U_{{e}
3}|^{2}$ is small. Thus we see from Eq. (\ref{078}) that in the
framework of three-neutrino mixing the survival probability of the
solar $\nu_{e}$ (up to corrections not larger than a few \% ) has
the two neutrino form
\begin{equation}
\overline P(\nu_{e}\to\nu_{e}) \simeq
P^{(1,2)}(\nu_{e}\to\nu_{e})
\label{080}
\end{equation}
and is characterized by two parameters  $\Delta m^{2}_{21}= \Delta
m^{2}_{\rm{sol}}$ and $\tan \,\theta_{12}=\tan
\,\theta_{\rm{sol}}$. Let us stress  that the reasons for
this aproximate decoupling \cite{BG} of the oscillations of solar neutrinos
from  oscillations of neutrinos in atmospheric and LBL experiments
are the hierarchy of the neutrino mass-squared differences
 and the smallness of $|U_{{e} 3}|^{2}$.

If the parameters $\Delta m^{2}_{\rm{sol}}$ and
$\tan \,\theta_{\rm{sol}}$ are in the MSW LMA region,
 the solar range of neutrino mass-squared differences
can be reached in reactor experiments. Such an experiment is
KamLAND \cite{KamLAND}. In this experiment $\bar \nu_{e}$'s from
several reactors in Japan at an average distance of about 170 km from
the detector will be recorded via the observation of the process

$$\bar \nu_{e}+ p \to e^{+}+ n. $$ The KamLAND detector is a tank
filled with a liquid scintillator (1000$ m^{3}$) and covered by
PMT's. About 700 events/kt/year are expected. After three years of
running the whole LMA region will be investigated in this
experiment.
\section{Conclusion}

In the last years significant progress in the investigation of the
phenomenon of neutrino oscillations, envisaged by B.Pontecorvo
many years ago, has been reached. The significant $\cos
\theta_{z}$-asymmetry of the Multi-GeV muon atmospheric neutrino
events, observed by the Super-Kamiokande collaboration,
constitutes convincing evidence for neutrino oscillations driven
by the neutrino mixing.

The comparison of the solar neutrino event rates, measured in the
Super-Kamiokande and in the SNO experiments, allowed to conclude
in a model independent way that solar $\nu_{e}$ on the way from
the sun to the earth  transfer into $\nu_{\mu}$ and $\nu_{\tau}$.

There exist, however, several fundamental questions to which
future experiments will be addressed:

\begin{enumerate}
\item
How many massive neutrinos exist in nature?

The minimal number that corresponds to three flavor neutrinos is
equal to three. If, however, the LSND result \cite{LSND} will be confirmed,
 in order to describe all neutrino oscillation data we
need three different neutrino mass-squared differences and (at
least) four massive neutrinos. The experiment MiniBooNE
\cite{MiniB}, started in 2002, plans to check the LSND claim in about two years.
\item

What is the nature of the massive neutrinos. Are they Dirac or Majorana particles?

The answer to this question can be obtained from experiments on
the search for neutrinoless double $\beta $ decay. The best lower
bound for the lifetime of this process was obtained in the
$^{76}$Ge Heidelberg-Moscow experiment \cite{HM} ($T_{1/2}\geq 1.9
\cdot 10^{25}y; \,~~90 \%CL$). From this result for the effective
 Majorana mass $|<m>| =| \sum_{i}U_{ei}^{2}\,~m_{i}|$ the bound
\begin{equation}
|<m>|\leq (0.2-0.6)\,~
\rm{eV}
\label{081}
\end{equation}
 can be  obtained.\footnote{ The bound on $|<m>|$ depends
on the value of the corresponding nuclear matrix element; the bound (\ref{081})
takes into account different calculations.}
Future experiments on the search for
neutrinoless double $\beta$ decay \cite{futdoubl} will have a
sensitivity $|<m>|\simeq 10^{-2}\rm{eV}$.

\item
What is the value of the minimal neutrino mass $m_{1}$?

In the experiments on the investigation of the effects of
neutrino masses  by the measurement of the high energy part of the
 $\beta $ spectrum of  $^3$H it was found that 
 $m_{1}\leq 2.2\,\rm{eV} $ \cite{Mainz} and  $m_{1}\leq
 2.5\,\rm{eV}\, $ \cite{Troitsk}.
In the future KATRIN experiment \cite{Katrin} a sensitivity
$m_{1}\simeq (0.3-0.4)\,~\rm{eV}$ is planned to be
reached.

\end{enumerate}

The precise measurement of the parameters $\Delta m^{2}_{31}$,
$\theta_{23}$, $|U_{e 3}|^{2}$, the investigation of the character
of the neutrino mass spectrum (hierarchy or reversed hierarchy),
the search for effects of CP violation in the lepton sector etc.
constitute the programs for long baseline experiments with
neutrinos from Super Beam facilities \cite{jhf} and Neutrino
Factories \cite{Cline}.

The small neutrino masses could be a signature of a new large
scale at which the lepton number is violated \cite{see-saw}. There
exist, however, other possible explanations for the smallness of
neutrino masses (e.g., large extra dimensions\cite{largedim} and
others).

It is obvious that a lot of new experimental and theoretical
efforts are required in order to reveal the {\em true origin of the
newly discovered phenomenon of small neutrino masses and neutrino
mixing}.

It is a pleasure for me to thank W.Potzel for valuable suggestions.

\end{document}